\documentclass[notitlepage,twocolumn,prb,aps,showpacs,10pt,superscriptaddress]{revtex4-1}
\usepackage{amsmath}
\usepackage{graphicx}
\newcommand{\mbf}[1]{\mathbf{#1}}
\renewcommand{\t}[1]{\textrm{#1}}
\newcommand{\nn}{\nonumber\\}
\newcommand{\an}{\allowdisplaybreaks\\}
\renewcommand{\k}{\mbf k}
\renewcommand{\Im}{\textrm{Im}}
\renewcommand{\Re}{\textrm{Re}}

\newcommand{\up}{\uparrow}
\newcommand{\down}{\downarrow}
\newcommand{\Jsd}{J_{\textrm{sd}}}
\newcommand{\Jpd}{J_{\textrm{pd}}}
\newcommand{\Hsd}{H_{\textrm{sd}}}
\newcommand{\Himp}{H_{\textrm{imp}}}
\newcommand{\HZe}{H_Z^{\textrm{e}}}
\newcommand{\HZMn}{H_Z^{\textrm{Mn}}}
\newcommand{\we}{\omega_{\textrm{e}}}
\newcommand{\wMn}{\omega_{\textrm{Mn}}}
\newcommand{\bwe}{\boldsymbol\omega_{\textrm{e}}}
\newcommand{\bwMn}{\boldsymbol\omega_{\textrm{Mn}}}

\newcommand{\NMn}{N_{\textrm{Mn}}}
\newcommand{\Nimp}{N_{\textrm{imp}}}
\newcommand{\bstd}{{b_{\sigma_1 n_1 \k_1}^{\sigma_2n_2\k_2}}}
\newcommand{\cstd}{{c_{\sigma_1 \k_1}^{\sigma_2\k_2}}}
\newcommand{\ddt}[0]{\frac{\partial}{\partial t}}
\begin{document}
\title{Influence of non-magnetic impurity scattering on the
spin dynamics in diluted magnetic semiconductors}
\author{M.~Cygorek}
\affiliation{Theoretische Physik III, Universit{\"a}t Bayreuth, 95440 Bayreuth, Germany}
\author{F.~Ungar}
\affiliation{Theoretische Physik III, Universit{\"a}t Bayreuth, 95440 Bayreuth, Germany}
\author{P.~I.~Tamborenea}
\affiliation{Departamento de F\'isica and IFIBA, FCEN, Universidad de Buenos Aires, Ciudad
Universitaria, Pabell\'on I, 1428 Ciudad de Buenos Aires, Argentina }
\author{V.~M.~Axt}
\affiliation{Theoretische Physik III, Universit{\"a}t Bayreuth, 95440 Bayreuth, Germany}
\begin{abstract}
The doping of semiconductors with magnetic impurities gives rise not only to
a spin-spin interaction between quasi-free carriers and magnetic impurities,
but also to a local spin-independent disorder potential for the carriers.
Based on a quantum kinetic theory for the carrier and impurity density
matrices as well as the magnetic and non-magnetic carrier-impurity 
correlations, the influence of the non-magnetic scattering potential on the 
spin dynamics in DMS after optical excitation with circularly polarized light
is investigated using the example of Mn-doped CdTe. 
It is shown that non-Markovian effects, which are predicted in 
calculations where only the magnetic carrier-impurity interaction is 
accounted for, can be strongly suppressed in the presence of non-magnetic
impurity scattering. 
This effect can be traced back to a significant
redistribution of carriers in $\k$-space which is enabled by
the build-up of large carrier-impurity correlation energies. 
A comparison with the Markov limit of the quantum kinetic theory shows 
that, in the presence of an external magnetic field parallel to the
initial carrier polarization, the asymptotic value
of the spin polarization at long times is significantly different in the
quantum kinetic and the Markovian calculations. This effect can also be
attributed to the formation of strong correlations which invalidates 
the semiclassical Markovian picture and it is stronger when 
the non-magnetic carrier-impurity interaction is accounted for.
In an external magnetic field perpendicular to the initial carrier spin,
the correlations are also responsible for a renormalization of the
carrier spin precession frequency. 
Considering only the magnetic carrier-impurity interaction,
a significant renormalization is predicted for a very limited set of
material parameters and excitation conditions. Accounting
also for the non-magnetic interaction a relevant renormalization of the
precession frequency is found to be more ubiquitous.

\end{abstract}
\pacs{75.78.Jp, 75.50.Pp, 75.30.Hx, 72.10.Fk}
\maketitle
\section{Introduction}
Most of the devices based on the spintronics paradigm that are commercially
available today use the fact that spin-up and spin-down carriers 
exhibit different transmission and reflection probabilities 
at interfaces involving ferromagnetic metals \cite{GMR1,Spintronics}.
However, some applications, like spin transistors \cite{DattaDas}, 
require the control not only of spin-up and spin-down occupations,
but also of the coherent precession of spins perpendicular to the
quantization axis provided by the structure. For this purpose,
spintronic devices based on semiconductors are preferable to metallic
structures since the dephasing time in a metal is about three
orders of magnitude shorter than in a semiconductor \cite{Awschalom07}.
In the context of semiconductor spintronics 
\cite{Zutic,spintronics_dietl,spintronics_ohno},
a particularly interesting class of materials for future applications
are diluted magnetic semiconductors (DMS) \cite{Dietl14,
DOM,BenCheikh2013,PulsedB_DMS,HyperfineCdMnTe,Debus2016,Debus2014,Debus2010,
SpinWavesPerezCibert11,Perakis_Wang,perakis08,Krenn,Crooker97,Barate10,Wu09},
which are obtained when semiconductors are doped with transition metal
elements, such as Mn, which act as localized magnetic moments.
While some types of DMS, such as Ga$_{1-x}$Mn$_x$As, 
exhibit a ferromagnetic phase
\cite{OhnoFerromagGaMnAs,Dietl14}, 
other types of DMS, like the usually paramagnetic CdMnTe, are 
especially valued for the enhancement of the effective carrier g-factor
by the giant Zeeman effect that can be used, e.g., to facilitate
an injection of a spin-polarized current into a light-emitting diode
\cite{Fiederling}.
Besides causing the giant Zeeman effect, the 
$s$-$d$ exchange interaction between the quasi-free carriers and localized
magnetic impurities also leads to other effects, such as inducing 
spin-flip scattering and thereby a direct transfer of spins from
the carriers to the impurities and vice versa 
\cite{Thurn:13_1,Cygorek:14_1,PESC,kdep}.

Typically, the $s$-$d$ interaction is described by a Kondo-like
\cite{Kondo} localized spin-spin interaction between carriers and
impurities. However, in real DMS materials, the introduction of Mn impurities
not only leads to a spin-dependent interaction Hamiltonian, but also
to a spin-independent local potential for the carriers 
\cite{KossutRate3D}. The reason for the appearance of 
this spin-independent potiential is that, in the case of Cd$_{1-x}$Mn$_x$Te,
the semiconductor CdTe has a different band structure than MnTe and
carriers located at unit cells with Mn impurities experience a larger
local potiential energy than carriers at unit cells with Cd cations.
The strength of this local potential can be estimated by the conduction and
valence band offsets between CdTe and MnTe. Note, however, that usually, 
CdTe crystallizes in a zinc-blende structure, while MnTe is found in 
a wurzite structure. Thus, a better estimation for the strength of the
spin-independent local potential is obtained by studying 
CdTe/Cd$_{1-x}$Mn$_x$Te heterostructures where both materials 
appear in the form of a zinc-blende lattice \cite{CdTe_offset_Nawrocki}.
From such investigations, the strenght of the local spin-independent potential 
for carriers at Mn sites of about 1.6 eV can be estimated.
In contrast, the spin-dependent local interaction in DMS is typically about
220 meV, i.e. one order of magnitude lower.
This consideration suggests that the non-magnetic impurity scattering 
caused by the local spin-independent interaction between carriers and impurities
should not be disregarded in the study of the spin physics in DMS.

It is noteworthy that a theory which takes into account a local magnetic 
interaction as well as a non-magnetic local potential in a DMS, 
the V-J tight-binding model was employed to study the magnetic properties 
of GaMnAs \cite{Bouzerar_Unified}
and it was found that taking into account 
the non-magnetic interaction is necessary in order to obtain 
results in good quantitative agreement with 
\emph{ab initio} calculations for the Curie temperature and with
experimental data for the optical conductivity.

For the spin dynamics, scattering at non-magnetic impurities has 
already important consequences in non-magnetic semiconductors \cite{WuReview}
in the presence of spin-orbit fields, 
where scattering processes can enhance or reduce the
spin relaxation and dephasing significantly, e.g., via the Elliott-Yafet
\cite{EY2} and D'yakonov-Perel' \cite{DP} mechanisms.

The goal of the present article is to investigate how
the non-magnetic interaction between carriers and impurities
affects the spin dynamics in 
paramagnetic II-VI DMS.
To this end we employ a quantum kinetic theory for carrier and impurity
density matrices including the carrier-impurity correlations
starting from a system Hamiltonian that comprises 
a kinetic energy term, 
the magnetic and non-magnetic carrier-impurity interactions as well as
the carrier and impurity Zeeman energies.
Earlier quantum kinetic studies of the spin dynamics in DMS
\cite{Thurn:12,Thurn:13_1,Thurn:13_2,Cygorek:14_1,PESC},
which only considered the spin-dependent $s$-$d$ interaction, 
predicted that in some cases, such as in narrow quantum wells optically excited
very close to the band edge \cite{Proceedings}, 
the spin transfer between carriers and impurities
cannot be well described by rate equations. 
Rather, the time evolution of the carrier spin is, in these cases,
non-exponential and it can exhibit non-monotonic features such as overshoots.
These effects are non-Markovian, as they can be traced back to the
finite memory provided by the correlations, since the
Markovian assumption of a $\delta$-like memory leads to effective rate 
equations that predict an exponential spin dynamics\cite{kdep}.

Here, we find that these non-Markovian effects predicted in the theory
of Refs.~\onlinecite{Thurn:12,Thurn:13_1,Thurn:13_2,Cygorek:14_1,PESC}
are suppressed in the case of the conduction band of a 
Cd$_{1-x}$Mn$_x$Te quantum well when non-magnetic scattering 
of carriers at the impurities is taken into account. 
While, in this case, the non-monotonic behavior of the spin dynamics 
disappears, the quantum kinetic theory predicts quantitative changes in the
effective spin transfer rate compared with the Fermi's golden rule value.
The suppression of the non-Markovian features
is mainly caused by a significant redistribution of carriers
away from the band edge where the non-Markovian effects are particularly 
strong \cite{Proceedings}. This carrier redistribution
is facilitated by the build-up of strong 
carrier-impurity correlations providing a correlation energy of the order
of a few meV per electron that leads to an increase of the average 
kinetic electron energy by about the same amount.
Due to the different strengths of the interactions 
in the conduction band of Cd$_{1-x}$Mn$_x$Te, 
the non-magnetic carrier-impurity correlation energy is also much larger than 
the magnetic correlation energy studied before in Ref.~\onlinecite{FreqRenorm,*FreqRenorm_Erratum}.

In other cases, such as in the
valence band of Cd$_{1-x}$Mn$_x$Te, the non-magnetic impurty scattering
can be much weaker than the magnetic spin-flip scattering and
the non-Markovian effects prevail.

In the presence of an external magnetic field parallel to the initial
carrier spin polarization, it was shown \cite{SPIE}
that a quantum kinetic treatment of the magnetic part of the 
carrier-impurity interaction in DMS leads to a significantly different
asymptotic value of the carrier spin polarization at long times $t$.
Because this is also a consequence of an energetic redistribution of 
carriers, including non-magnetic scattering increases this effect.
If the initial carrier spin polarization is perpendicular to the
external magnetic field, the carrier spins precess about the effective
field comprised of the external field and the mean field due to the impurity
magnetization. As shown in Ref.~\onlinecite{FreqRenorm,*FreqRenorm_Erratum}, the 
carrier-impurity correlations built up by the magnetic $s$-$d$ interaction
renormalize the carrier spin precession frequency. Here, we show that
when both, the magnetic and the non-magnetic interactions are taken
into account the renormalization of the carrier spin precession frequency
can be different in sign and magnitude compared with calculations 
in which only the magnetic interaction is considered. 

The article is structured as follows:
First, quantum kinetic equations of motion for the carrier and impurity
density matrices as well as for the magnetic and non-magnetic carrier-impurity 
correlations are formulated for a DMS with magnetic and non-magnetic 
carrier-impurity interactions. Then, we derive the Markov limit of the 
quantum kinetic theory which enables a comparison and allows us to 
distinguish the genuine quantum kinetic effects
from the Markovian behavior. Furthermore, from the Markov
limit we can derive analytic expressions for the carrier-impurity correlation
energies as well as the correlation-induced renormalization of the
carrier spin precession frequency.
After having layed out the theory, we present numerical simulations
of the quantum kinetic equations for the conduction band of a 
Cd$_{1-x}$Mn$_x$Te quantum well 
including magnetic and non-magnetic scattering at the Mn impurities
and discuss the energetic 
redistribution of carriers as well as the correlation energies.
Then, we estimate the influence of non-magnetic impurity interaction 
on the spin dynamics in the valence band of Cd$_{1-x}$Mn$_x$Te.
Finally, we discuss the effects of the
non-magnetic impurity scattering on the spin dynamics in DMS in the 
presence of an external magnetic field parallel and perpendicular to
an initial non-equilibrium carrier spin polarization.

\section{Theory}
\subsection{DMS Hamiltonian}
Here, we consider an intrinsic DMS such as Cd$_{1-x}$Mn$_x$Te 
in the presence of an external magnetic field.
The total Hamiltonian of this DMS is given by
\begin{subequations}
\label{eq:Htot}
\begin{align}
H=&H_0+\Hsd+\Himp+\HZe+\HZMn,\\
H_0=&\sum_{\k\sigma} \hbar\omega_\k c^\dagger_{\sigma\k}c_{\sigma\k},\\
\Hsd=&\frac{\Jsd}{V}\sum_{\k\k'\sigma\sigma'}\sum_{Inn'}
\mbf S_{nn'}\cdot\mbf s_{\sigma\sigma'} 
c^\dagger_{\sigma\k}c_{\sigma'\k'}e^{i(\k'-\k)\mbf R_{I}} 
\hat{P}^I_{nn'},\\
\Himp=&\frac{J_0}{V}\sum_{\k\k'\sigma}\sum_J
c^\dagger_{\sigma\k}c_{\sigma\k'}e^{i(\k'-\k)\mbf R_{J}}, \\
H_Z^{\t{e}}=&\sum_{\k\sigma\sigma'}\hbar g_e\mu_B \mbf B\cdot \mbf s_{\sigma\sigma'}
c^\dagger_{\sigma\k}c_{\sigma'\k},\\
H_Z^{\t{Mn}}=&\sum_{Inn'}\hbar g_\t{Mn}\mu_B\mbf B\cdot\mbf S_{nn'}
\hat{P}^I_{nn'},
\end{align}
\end{subequations}
where $H_0$ is the single-electron Hamiltonian due to the crystal potential,
$H_{\t{sd}}$ describes the magnetic $s$-$d$ exchange interaction between
the carriers and the impurities, $H_{\t{imp}}$ describes the spin-independent
scattering of carriers at impurities and $H_Z^{\t{e}}$ and $H_Z^{\t{Mn}}$ are the
carrier and impurity Zeeman energies.

In Eqs.~(\ref{eq:Htot}), $c^\dagger_{\sigma\k}$ and $c_{\sigma\k}$ 
denote the 
creation and annihilation operators for conduction band electrons
with wave vector $\k$ in the spin subband 
$\sigma=\{\up,\down\}$. 
The magnetic Mn impurities are described by the operator
$\hat{P}^I_{nn'}=|I,n\rangle\langle I,n'|$ where $|I,n\rangle$ is
the $n$-th spin state ($n\in\{-\frac 52,-\frac 32,\dots \frac 52\}$)
of the $I$-th magnetic impurity located at $\mbf R_I$.
The band structure of the semiconductor is described by
$\hbar\omega_\k$, which we assume to be parabolic 
$\omega_{\k}=\frac{\hbar\k^2}{2m^*}$
with effective mass $m^*$. $V$ denotes the volume of the sample.
$\Jsd$ is the $s$-$d$ coupling constant for the spin-spin interaction
between carriers and impurities and
$J_0$ is the non-magnetic coupling constant.
$\mbf S_{n_1n_2}$ and $\mbf s_{\sigma_1\sigma_2}$ are the vectors with
components consisting of spin-$\frac 52$ and spin-$\frac 12$ 
spin matrices for the impurities and
the conduction band electrons, respectively, where the
unit $\hbar$ has been substituted into the definition of $\Jsd$ so that
$\mbf s_{\sigma_1\sigma_2}=\frac 12\boldsymbol\sigma_{\sigma_1\sigma_2}$,
where $\boldsymbol\sigma_{\sigma_1\sigma_2}$ are the Pauli matrices.
Finally, $g_e$ and $g_\textrm{Mn}$ are the g-factors of the electrons and the
impurities, respectively, and $\mu_B$ is the Bohr magneton.

In order to account for spin-independent scattering not only at 
Mn impurities but also additional non-magnetic scattering centers,
such as in quaternary compound DMSs like HgCdMnTe \cite{Ungar15},
we allow the number of scattering centers $\Nimp$ in general 
to be larger than the number of magnetic impurities 
$\NMn$. Here, we use the notation that the index $I$ runs from 
$1$ to $\NMn$ while the index $J$ runs from $1$ to $\Nimp$. 

\subsection{Quantum kinetic equations of motion}
The goal of this artilce is to study 
the time evolution of the carrier spin polarization
after optical excitation with circularly polarized light
which can be extracted from the carrier density matrix.
In this section, we derive the corresponding equations of motion
starting from the total Hamiltonian in Eqs.~(\ref{eq:Htot}).

Following Ref.~\onlinecite{Thurn:12}, where for the conduction band
only $H_0$ and $\Hsd$ in Eqs.~(\ref{eq:Htot}) were considered, 
we seek to obtain a closed set of equations 
for the reduced carrier and impurity density matrices as well as
for the carrier-impurity correlations:
\begin{subequations}
\label{eq:dynvars}
\begin{align}
M_{n_1}^{n_2}=&\langle \hat{P}^I_{n_1n_2}\rangle\\
C_{\sigma_1\k_1}^{\sigma_2}=&\langle c^\dagger_{\sigma_1\k_1}c_{\sigma_2\k_1}
\rangle, \\
\bar{C}_{\sigma_1\k_1}^{\sigma_2\k_2}=
&V\langle c^\dagger_{\sigma_1\k_1}c_{\sigma_2\k_2}e^{i(\k_2-\k_1)\mbf R_J}
\rangle, &\textrm{for }\k_2\neq\k_1,\\
Q_{\sigma_1n_1\k_1}^{\sigma_2n_2\k_2}=
&V\langle c^\dagger_{\sigma_1\k_1}c_{\sigma_2\k_2}e^{i(\k_2-\k_1)\mbf R_I}
\hat{P}^I_{n_1n_2}\rangle, &\textrm{for }\k_2\neq\k_1.
\end{align}
\end{subequations}
$M_{n_1}^{n_2}$ and $C_{\sigma_1\k_1}^{\sigma_2}$ are the 
impurity and electron density matrices and
$\bar{C}_{\sigma_1\k_1}^{\sigma_2\k_2}$ as well as 
$Q_{\sigma_1n_1\k_1}^{\sigma_2n_2\k_2}$ are the non-magnetic and magnetic
carrier-impurity correlations, respectively.
In Eqs.~(\ref{eq:dynvars}), the brackets denote not only the 
quantum mechanical average of the operators,
but also an average over a random distribution of impurity 
positions, which we assume to be on average homogeneous so that 
$\langle e^{i(\k_2-\k_1)\mbf R_J}\rangle=\delta_{\k_1\k_2}$.

The equations of motion for the variables defined in Eqs.~(\ref{eq:dynvars})
can be derived using the Heisenberg equations of motion for the
corresponding operators. Note, however, that this procedure leads
to an infinite hierarchy of variables and equations of motion, since, e.~g.,
the equation of motion for 
$\langle c^\dagger_{\sigma_1\k_1}c_{\sigma_2\k_2}e^{i(\k_2-\k_1)\mbf R_I}
\hat{P}^I_{n_1n_2}\rangle$ contains also terms of the form 
$\langle c^\dagger_{\sigma_1\k_1}c_{\sigma\k}e^{i(\k-\k_1)\mbf R_I}
e^{i(\k_2-\k)\mbf R_{I'}}\hat{P}^I_{n_1n_2}\hat{P}^{I'}_{nn'}\rangle$
for $I'\neq I$ which cannot be expressed in terms of the variables in
Eqs.~(\ref{eq:dynvars}). Thus, in order to obtain a closed set of equations,
one has to employ a truncation scheme. Here, we follow the procedure of 
Ref.~\onlinecite{Thurn:12}: we factorize the averages over products of 
operators and define the true correlations to be the remainder when all 
combinations of factorizations have been subtracted from the averages.
For example, we define (for $\k_2\neq\k_1$)
\begin{align}
&\delta\langle c^\dagger_{\sigma_1\k_1}c_{\sigma_2\k_2}e^{i(\k_2-\k_1)\mbf R_I}
\hat{P}^I_{n_1n_2}\rangle:=\nn&
\langle c^\dagger_{\sigma_1\k_1}c_{\sigma_2\k_2}e^{i(\k_2-\k_1)\mbf R_I}
\hat{P}^I_{n_1n_2}\rangle \nn&-\Big(
\langle c^\dagger_{\sigma_1\k_1}c_{\sigma_2\k_2}\rangle 
\langle e^{i(\k_2-\k_1)\mbf R_I}\rangle\langle \hat{P}^I_{n_1n_2}\rangle \nn&
+\langle c^\dagger_{\sigma_1\k_1}c_{\sigma_2\k_2}
e^{i(\k_2-\k_1)\mbf R_I}\rangle\langle \hat{P}^I_{n_1n_2}\rangle \nn&
+\langle e^{i(\k_2-\k_1)\mbf R_I}\rangle
\langle c^\dagger_{\sigma_1\k_1}c_{\sigma_2\k_2}\hat{P}^I_{n_1n_2}\rangle
\Big)
\end{align}
where $\delta\langle \dots\rangle$ denotes the true correlations. 
The basic assumption of the truncation scheme of Ref.~\onlinecite{Thurn:12} is
that all correlations higher than 
$\delta\langle c^\dagger_{\sigma_1\k_1}c_{\sigma_2\k_2}
e^{i(\k_2-\k_1)\mbf R_I}\rangle$ and 
$\delta\langle c^\dagger_{\sigma_1\k_1}c_{\sigma_2\k_2}
e^{i(\k_2-\k_1)\mbf R_I}\hat{P}^I_{n_1n_2}\rangle$ are negligible. 
This assumption results in a closed set of equations of motion for the
reduced density matrices and the true correlations. 
However, it turns out \cite{Cygorek:14_1} that the equations of motion 
can be written down in a more condensed form when switching back to
the full (non-factorized) higher order density matrices as variables, after the
higher (true) correlations are neglected. For details of this procedure,
the reader is referred to Refs.~\onlinecite{Thurn:12,Cygorek:14_1}.

Applying this truncation scheme to the total Hamiltonian (\ref{eq:Htot}) 
including magnetic and non-magnetic carrier-impurity interactions as well 
as the Zeeman terms for carriers and impurities leads to the 
equations of motion for the variables defined in Eqs.~(\ref{eq:dynvars}):
\begin{widetext}
\begin{subequations}
\begin{align}
-i\hbar\ddt M_{n_1}^{n_2}=&
\sum_n\hbar\boldsymbol\wMn\cdot
(\mbf S_{nn_1}M_{n}^{n_2}-\mbf S_{n_2n}M_{n_1}^{n})
+\frac{\Jsd}{V^2}\sum_n\sum_{\k\k'\sigma\sigma'}
(\mbf S_{nn_1}\cdot \mbf s_{\sigma\sigma'}Q_{\sigma n\k}^{\sigma' n_2\k'}-
\mbf S_{n_2n}\cdot\mbf s_{\sigma\sigma'}Q_{\sigma n_1\k}^{\sigma' n\k'})\big],\an
-i\hbar\ddt C_{\sigma_1\k_1}^{\sigma_2}=&
\sum_{\sigma}\hbar\boldsymbol\omega_{e}\cdot
(\mbf s_{\sigma\sigma_1}C_{\sigma\k_1}^{\sigma_2}
-\mbf s_{\sigma_2\sigma}C_{\sigma_1\k_1}^{\sigma})
+\Jsd\frac{\NMn}{V^2}\sum_{nn'} \sum_{\k\sigma}
(\mbf S_{nn'}\cdot\mbf s_{\sigma\sigma_1}
Q_{\sigma n\k}^{\sigma_2n'\k_1}
-\mbf S_{nn'}\cdot\mbf s_{\sigma_2\sigma}Q_{\sigma_1 n\k_1}^{\sigma n'\k})
+\nn&+
J_0\frac{\Nimp}{V^2}\sum_\k (\bar{C}_{\sigma_1\k}^{\sigma_2\k_1}
-\bar{C}_{\sigma_1\k_1}^{\sigma_2\k}),\an
-i\hbar\ddt{Q}_{\sigma_1n_1\k_1}^{\sigma_2n_2\k_2}=&
\hbar(\omega_{\k_1}-\omega_{\k_2}){Q}_{\sigma_1n_1\k_1}^{\sigma_2n_2\k_2}
+{b_{\sigma_1 n_1\k_1}^{\sigma_2n_2\k_2}}^{I}
+{b_{\sigma_1 n_1\k_1}^{\sigma_2n_2\k_2}}^{II}
+{b_{\sigma_1 n_1\k_1}^{\sigma_2n_2\k_2}}^{III}
+{b_{\sigma_1 n_1\k_1}^{\sigma_2n_2\k_2}}^{\t{imp}},\an
-i\hbar\ddt\bar{C}_{\sigma_1\k_1}^{\sigma_2\k_2}=& 
\hbar(\omega_{\k_1}-\omega_{\k_2})\bar{C}_{\sigma_1\k_1}^{\sigma_2\k_2}
+{c_{\sigma_1\k_1}^{\sigma_2\k_2}}^{I}
+{c_{\sigma_1\k_1}^{\sigma_2\k_2}}^{II}
+{c_{\sigma_1\k_1}^{\sigma_2\k_2}}^{III}
+{c_{\sigma_1\k_1}^{\sigma_2\k_2}}^{\t{sd}}
\end{align}
with
\begin{align}
{b_{\sigma_1 n_1\k_1}^{\sigma_2n_2\k_2}}^{I}=&\sum_{n\sigma\sigma'}
\Jsd[\mbf S_{nn_1}\cdot\mbf s_{\sigma\sigma'}(\delta_{\sigma_1\sigma'}
-C_{\sigma_1\k_1}^{\sigma'})C_{\sigma\k_2}^{\sigma_2}M_{n}^{n_2}-
\mbf S_{n_2n}\cdot\mbf s_{\sigma\sigma'}(\delta_{\sigma\sigma_2}
-C_{\sigma\k_2}^{\sigma_2})C_{\sigma_1\k_1}^{\sigma'}M_{n_1}^{n}],\\
{b_{\sigma_1 n_1\k_1}^{\sigma_2n_2\k_2}}^{II}=&
\sum_\sigma\hbar\boldsymbol\omega_e\cdot
(\mbf s_{\sigma\sigma_1}Q_{\sigma n_1\k_1}^{\sigma_2 n_2\k_2}-
\mbf s_{\sigma_2\sigma}Q_{\sigma_1 n_1\k_1}^{\sigma n_2\k_2})
+\sum_n \hbar\boldsymbol\wMn\cdot
(\mbf S_{nn_1}Q_{\sigma_1n\k_1}^{\sigma_2n_2\k_2}
-\mbf S_{n_2n}Q_{\sigma_1n_1\k_1}^{\sigma_2n\k_2}),\\
{b_{\sigma_1 n_1\k_1}^{\sigma_2n_2\k_2}}^{III}=&
\frac{\Jsd}V\sum_n\sum_{\k\sigma} \Big\{
(\mbf S_{nn_1}\cdot\mbf s_{\sigma\sigma_1}Q_{\sigma n\k}^{\sigma_2n_2\k_2}
-\mbf S_{n_2n}\cdot\mbf s_{\sigma_2\sigma}Q_{\sigma_1n_1\k_1}^{\sigma n\k})\nn&
-\sum_{\sigma'}\mbf s_{\sigma\sigma'}\cdot
\big[ C_{\sigma_1\k_1}^{\sigma'} \big(
\mbf S_{nn_1}Q_{\sigma n\k}^{\sigma_2 n_2 \k_2}
-\mbf S_{n_2n}Q_{\sigma n_1\k}^{\sigma_2 n\k_2}\big)
+C_{\sigma\k_2}^{\sigma_2}\big(
\mbf S_{nn_1}Q_{\sigma_1n\k_1}^{\sigma'n_2\k}
-\mbf S_{n_2n}Q_{\sigma_1n_1\k_1}^{\sigma'n\k}\big)\big]\Big\},\\
{b_{\sigma_1 n_1\k_1}^{\sigma_2n_2\k_2}}^{\t{imp}}=&
J_0 \big[\big(C_{\sigma_1\k_2}^{\sigma_2}-
C_{\sigma_1\k_1}^{\sigma_2}\big)M_{n_1}^{n_2}
+\frac 1V\sum_{\k} \big({Q}_{\sigma_1n_1\k}^{\sigma_2n_2\k_2}
-Q_{\sigma_1n_1\k_1}^{\sigma_2n_2\k}\big)\big],
\end{align}
and
\begin{align}
{c_{\sigma_1\k_1}^{\sigma_2\k_2}}^{I}=&
J_0(C_{\sigma_1\k_2}^{\sigma_2}-C_{\sigma_1\k_1}^{\sigma_2}),\\
{c_{\sigma_1\k_1}^{\sigma_2\k_2}}^{II}=&
\sum_\sigma \hbar\boldsymbol\omega_e\cdot(
\mbf s_{\sigma\sigma_1}\bar{C}_{\sigma\k_1}^{\sigma_2\k_2}
-\mbf s_{\sigma_2\sigma}\bar{C}_{\sigma_1\k_1}^{\sigma\k_2}),\\
{c_{\sigma_1\k_1}^{\sigma_2\k_2}}^{III}=&
\frac{J_0}V\sum_\k(\bar{C}_{\sigma_1\k}^{\sigma_2\k_2}
-\bar{C}_{\sigma_1\k_1}^{\sigma_2\k}),\\
{c_{\sigma_1\k_1}^{\sigma_2\k_2}}^{\t{sd}}=&
\Jsd\sum_{nn'}\sum_\sigma M_{nn'}\mbf S_{nn'}\cdot
\big(\mbf s_{\sigma\sigma_1}C_{\sigma\k_2}^{\sigma_2}
-\mbf s_{\sigma_2\sigma}C_{\sigma_1\k_1}^{\sigma}\big)
+\frac{\Jsd}V\frac{\NMn}{\Nimp}\sum_{nn'}\sum_{\k\sigma}
\mbf S_{nn'}\cdot\big(\mbf s_{\sigma\sigma_1}Q_{\sigma n\k}^{\sigma_2 n'\k_2}
-\mbf s_{\sigma_2\sigma}Q_{\sigma_1 n\k_1}^{\sigma n'\k}\big),
\end{align}
\label{eq:corful}
\end{subequations}
\end{widetext}
where ${b_{\sigma_1 n_1\k_1}^{\sigma_2n_2\k_2}}^{X}$ are the source
terms for the magnetic carrier-impurity correlations, 
${c_{\sigma_1\k_1}^{\sigma_2\k_2}}^{X}$ are the sources for
the non-magnetic correlations and 
\begin{subequations}
\begin{align}
&\boldsymbol\wMn=g_{\t{Mn}}\mu_B\mbf B+
\frac{\Jsd}{\hbar}\frac 1V\sum_{\k\sigma\sigma'}
\mbf s_{\sigma\sigma'}C_{\sigma\k}^{\sigma'},\\
&\bwe=g_{\t{e}}\mu_B\mbf B+
\frac{\Jsd}{\hbar}\frac{\NMn}V\sum_{nn'}\mbf S_{nn'}M_{nn'}
\end{align}
\end{subequations}
are the mean-field precession frequencies of the impurity and carrier
spins, respectively. 
The first terms on the right-hand side
of Eqs.~(\ref{eq:corful}a) and (\ref{eq:corful}b)
represent the precession of the impurity and carrier spins in the 
mean field due to the carrier and impurity magnetization as well as the
external magnetic field.
The second terms in Eqs.~(\ref{eq:corful}a) and (\ref{eq:corful}b)
describe the effects of the magnetic carrier-impurity correlations on the 
impurity and carrier density matrices and the last term 
of Eq.~(\ref{eq:corful}b) describes the scattering of carriers at
non-magnetic impurities.

In analogy to the situation without non-magnetic impurity scattering ($J_0=0$) 
studied in Ref.~\onlinecite{Cygorek:14_1},
we label the source terms of the correlations on the right-hand side of 
the Eqs.~(\ref{eq:corful}c) and (\ref{eq:corful}d) as follows:
The terms $\bstd^I$ are the inhomogeneous driving terms depending only on 
single-particle quantities. $\bstd^{II}$ are homogeneous terms which
describe a precession-type motion of the correlations in the
effective fields $\bwe$ and  $\bwMn$.
The source terms $\bstd^{III}$ comprise the driving of the magnetic correlations
by other magnetic correlations with different wave vectors and 
describe a change of the wave vectors of the correlations due to the
$s$-$d$ interaction.
$\bstd^{\t{imp}}$ denotes the contributions to the equation for the 
magnetic correlations due to the non-magnetic impurity scattering.
The source terms $\cstd^X$ for the non-magnetic correlations are
classified analogously.

A straightforward but lengthy calculation confirms that Eqs.~(\ref{eq:corful})
conserve the particle number as well as the total energy
comprised of the single-particle contributions and the correlation energies.

\subsection{Markov limit}
Although Eqs.~(\ref{eq:corful}) can readily be used to calculate the 
spin dynamics given a set of appropriate initial conditions, 
it is instructive also to derive the Markov limit 
of the quantum kinetic equations \cite{Cygorek:14_1,PESC,kdep}.
On the one hand, this enables us to distinguish the Markovian behavior 
from genuine quantum kinetic effects. On the other hand, it allows us 
to derive analytic expressions for the correlation energies
and the renormalization of the precession frequencies in the 
presence of an external magnetic field \cite{FreqRenorm,*FreqRenorm_Erratum}.

The derivation of the Markov limit comprises two steps \cite{kdep}:
First, the equations of motion for the correlations are formally integrated
yielding explicit expressions for the correlations in the form of a
memory integral.
This yields integro-differential equations for the 
single-particle variables, where the values of the single-particle
variables at earlier times enter. 
Second, the memory integral is eliminated by assuming a $\delta$-like
short memory. 

However, the first step, which involves the formal integration of
the carrier-impurity correlations, can, in general, be complicated.
Nevertheless, if the source terms $\bstd^{III}$ and $\cstd^{III}$ 
as well as the correlation-dependent part of $\bstd^\t{imp}$ and
$\cstd^\t{sd}$ are neglected, the formal solution of 
Eqs.~(\ref{eq:corful}c-d) becomes much easier. In absence of 
non-magnetic impurity scattering, it has been shown that these source 
terms are indeed numerically insignificant \cite{Cygorek:14_1}. 
Furthermore, a straightforward calculation shows that neglecting these
terms also yields a consistent theory with respect to the conservation of the
total energy. Whether neglecting the terms $\bstd^{III}$, $\cstd^{III}$
and the correlation-dependent parts of $\bstd^\t{imp}$ and
$\cstd^\t{sd}$ is indeed a good approximation in the presence of
non-magnetic impurity scattering can be tested by comparing the 
numerical results of the quantum kinetic equations with and without 
accounting for these source terms.

Neglecting the aforementioned source terms in Eqs.~(\ref{eq:corful}), 
we first formulate a set of quantum kinetic equations 
for the new dynamical variables
\begin{subequations}
\begin{align}
&\langle S^i\rangle=\sum_{n_1n_2} S^i_{n_1n_2} M_{n_1n_2},\\
&n_\k=\sum_{\sigma} C_{\sigma\k}^{\sigma},\\
&s^i_\k=\sum_{\sigma_1\sigma_2} s^i_{\sigma_1\sigma_2}
C_{\sigma_1\k}^{\sigma_2},\\
&\bar{C}_{\phantom{\alpha}\k_1}^{\alpha\k_2}=
\sum_{\sigma_1\sigma_2} s^\alpha_{\sigma_1\sigma_2}
\bar C_{\sigma_1\k_1}^{\sigma_2\k_2}\\
&Q_{l\k_1}^{\alpha \k_2}=\sum_{\sigma_1\sigma_2} \sum_{n_1n_2}
s^\alpha_{\sigma_1\sigma_2}S^l_{n_1n_2}
Q_{\sigma_1n_1\k_1}^{\sigma_2n_2\k_2},
\end{align}
\label{eq:newvars}
\end{subequations}
where $\langle \mbf S\rangle$ is the average impurity spin and
$n_\k$ and $\mbf s_\k$ are the occupation density and spin density
of the carrier states with wave vector $\k$, respectively.
$\bar{C}_{\phantom{\alpha}\k_1}^{\alpha\k_2}$
as well as $Q_{l\k_1}^{\alpha \k_2}$ comprise the non-magnetic and 
magnetic carrier-impurity correlations. In Eqs.~(\ref{eq:newvars})
we use a notation in which
the Latin indices are in the range $\{1,2,3\}$, while the Greek indices
also include the value 0, where $s^0_{\sigma_1\sigma_2}=
\delta_{\sigma_1\sigma_2}$ is the 2x2 identity matrix.
The corresponding equations of motion for the variables
defined in Eqs.~(\ref{eq:newvars}) are explicitly given in 
appendix \ref{app:neweqs}.

Note that the source terms ${b_{l\k_1}^{\alpha\k_2}}^I$ for
the correlations $Q_{l\k_1}^{\alpha \k_2}$ depend on 
the second moments of the impurity spins
$\langle  S^i S^j\rangle=\sum\limits_{n_1n_2n_3}S^i_{n_1n_2}S^j_{n_2n_3}M_{n_1n_3}$
for which we do not present equations of motions, although such equations
can, in principle, be derived from Eqs.~(\ref{eq:corful}).
Here, we use the fact that for typical sample parameters the 
optically induced carrier density is usually much lower than the
impurity concentration, so that the average impurity spin only
changes marginally over time \cite{Cygorek:14_1}.
For the numerical calculations we assume that the
impurity density matrix can be approximately described 
as being in thermal equilibrium at all times where the
effective impurity spin temperature $T_{\t{Mn}}$ can be obtained
from the value of $\langle \mbf S\rangle$.
From this thermally occupied density matrix, 
the second moments $\langle  S^i S^j\rangle$ 
consistent with $\langle \mbf S\rangle$ can be calculated in each time 
step.

The equations of motion for the variables defined in Eqs.~(\ref{eq:newvars})
are the starting point for the formal integration
of the correlations. 
Note that Eqs.~(\ref{eq:eom_sum}d-g) for the correlations
$Q_{l\k_1}^{\alpha \k_2}$ and $\bar{C}_{\phantom{\alpha}\k_1}^{\alpha\k_2}$
can be transformed into the general form

\begin{align}
\label{eq:Qform}
\ddt Q_{\k_1}^{\k_2}&=-i(\omega_{\k_2}-\omega_{\k_1})Q_{\k_1}^{\k_2}
+i \chi_1\we Q_{\k_1}^{\k_2}\nn&
+i \chi_2\wMn Q_{\k_1}^{\k_2}
+{b_{\k_1}^{\k_2}}^I, 
\end{align}
where 
$\chi_1,\chi_2\in\{-1,0,1\}$
and the terms proportional to $\we=|\bwe|$ and $\wMn=|\bwMn|$ 
originate from the
precession of the correlations described by the source terms
$\bstd^{II}$ and $\cstd^{II}$. The term ${b_{\k_1}^{\k_2}}^I$ here denotes
the contributions from the source terms $\bstd^{I}$, $\cstd^I$,
$\bstd^\t{imp}$ and $\cstd^\t{sd}$ and only depends on the single-particle
variables.
The formal integration of Eq.~(\ref{eq:Qform}) yields
\begin{align}
\label{eq:Qint}
Q_{\k_1}^{\k_2}(t)=
\int\limits_0^t dt' 
e^{i[\omega_{\k_2}-(\omega_{\k_1}+\chi_1\we+\chi_2\wMn)](t'-t)}
{b_{\k_1}^{\k_2}}^I(t').
\end{align}

The Markov limit consists of assuming a short memory, i.e. the 
assumption that the correlations at time $t$ depend only significantly on
the single-particle variables at the same time $t$, so that one is 
inclined to evaluate ${b_{\k_1}^{\k_2}}^I(t')$ in Eq.~(\ref{eq:Qint}) at
$t'=t$ and to draw the source term out of the integral.
However, first, one has to make sure that the source terms are 
indeed slowly changing variables. For example, the carrier spin 
can precess rapidly about an external magnetic field. Therefore, we
first analyze the mean-field precession of the single-particle quantities
and split the source terms into parts oscillating with some frequencies
$\omega$ of the form
\begin{align}
\label{eq:bMF}
{b_{\k_1}^{\k_2}}^I(t')\stackrel{\t{MF}}=\sum_\omega\sum_{\chi\in\{-1,0,1\}} 
e^{i\chi\omega(t'-t)}{b_{\k_1}^{\k_2}}^{\omega,\chi}(t).
\end{align}
Then, the different oscillating parts ${b_{\k_1}^{\k_2}}^{\omega,\chi}(t)$
can be drawn out of the memory integral and the remaining 
integral can be solved in the limit of large times $t$ \cite{kdep}:
\begin{align}
\label{eq:memint}
\int\limits_0^t dt'\, e^{i\Delta \omega (t'-t)}
\stackrel{t\to\infty}{\longrightarrow}
\pi\delta(\Delta\omega)-\frac i{\Delta\omega}.
\end{align}

This procedure yields particularly transparent results in the case
where the external magnetic field and the impurity magnetization 
are collinear, as is usually the case when the number of impurities
exceeds the number of quasi-free carriers ($\NMn\gg N_e$), and the
impurity density matrix is initially occupied thermally.
Choosing the direction of $\bwe$ as a reference and defining 
$s_{\k_1}^\|:= \mbf s\cdot \frac{\bwe}\we$, 
$S^\|:=\hat{\mbf S}\cdot \frac{\bwe}\we$ and 
$\wMn^\|:=\bwMn\cdot\frac{\bwe}\we$,
the Markovian equations obtained for the 
spin-up and spin-down occupations and the perpendicular 
carrier spin component with respect to the direction of $\bwe$,

\begin{subequations}
\begin{align}
n^{\up/\down}_{\k_1}:=&\frac {n_{\k_1}}2 \pm s_{\k_1}^\|,\\
\mbf s^\perp_{\k_1}:=&\mbf s_{\k_1} - \frac{\bwe}\we s^\|_{\k_1},
\end{align}
\end{subequations}
are given by:
\begin{widetext}
\begin{subequations}
\label{eq:Markov}
\begin{align}
\ddt n_{\k_1}^{\up/\down}=&\frac{\pi}{\hbar^2 V^2}\sum_{\k_2}\bigg\{
\delta(\omega_{\k_2}-\omega_{\k_1})\big[
\Jsd^2\NMn\frac 12\langle {S^\|}^2\rangle 
\pm\Jsd J_0(\NMn+\Nimp)\langle S^\|\rangle
+2J_0^2\Nimp \big] 
(n^{\up/\down}_{\k_2}-n^{\up/\down}_{\k_1})+\nn&+
\delta\big[\omega_{\k_2}-\big(\omega_{\k_1}\pm (\we-\wMn^\|)\big)\big]
\Jsd^2\NMn\bigg[\Big(\langle {S^\perp}^2\rangle\pm\frac 12
\langle S^\|\rangle\Big)\big(1-n^{\up/\down}_{\k_1}\big)n_{\k_2}^{\down/\up}
-\Big(\langle {S^\perp}^2\rangle\mp\frac 12\langle S^\|\rangle\Big)
\big(1-n_{\k_2}^{\down/\up}\big)n^{\up/\down}_{\k_1}\bigg]\bigg\} ,\an
\ddt \mbf s_{\k_1}^\perp=&-\frac \pi{\hbar^2V^2}\sum_{\k_2}\bigg\{
\delta(\omega_{\k_2}-\omega_{\k_1})\Big[\Jsd^2\NMn\frac 12
\langle{S^\|}^2\rangle(\mbf s_{\k_2}^\perp+\mbf s_{\k_1}^\perp)
-2J_0^2\Nimp (\mbf s_{\k_2}^\perp-\mbf s_{\k_1}^\perp) \Big]\nn&
+\delta\big[\omega_{\k_2}-\big(\omega_{\k_1}+(\we-\wMn^\|)\big)\big]
\frac 12\Big(\langle {S^\perp}^2\rangle-\frac 12\langle S^\|\rangle
(1-2n_{\k_2}^\down)\Big) \mbf s_{\k_1}^\perp \nn&
+\delta\big[\omega_{\k_2}-\big(\omega_{\k_1}-(\we-\wMn^\|)\big)\big]
\frac 12\Big(\langle {S^\perp}^2\rangle+\frac 12\langle S^\|\rangle
(1-2n_{\k_2}^\up)\Big) \mbf s_{\k_1}^\perp \bigg\}\nn&
+\bwe\times \mbf s_{\k_1}^\perp +\frac 1{\hbar^2 V^2}\sum_{\k_2}\bigg\{
-\frac{\Jsd J_0}{\omega_{\k_2}-\omega_{\k_1}}\langle \mbf S\rangle\times
\big[ (\Nimp-\NMn)\mbf s_{\k_2}^\perp +(\NMn+\Nimp)\mbf s_{\k_1}^\perp\big]\nn&
-\frac{\Jsd^2\NMn}{\omega_{\k_2}-\big(\omega_{\k_1}+(\we-\wMn^\|)\big)}
\frac 12\Big(\langle {S^\perp}^2\rangle-\frac 12\langle S^\|\rangle
(1-2n_{\k_2}^\down)\Big)\Big(\frac{\bwe}{\we}\times \mbf s_{\k_1}^\perp\Big)\nn&
+\frac{\Jsd^2\NMn}{\omega_{\k_2}-\big(\omega_{\k_1}-(\we-\wMn^\|)\big)}
\frac 12\Big(\langle {S^\perp}^2\rangle+\frac 12\langle S^\|\rangle
(1-2n_{\k_2}^\up)\Big)\Big(\frac{\bwe}{\we}\times \mbf s_{\k_1}^\perp\Big)
\bigg\}.
\end{align}
\end{subequations}
\end{widetext}

The first line of the right-hand side of
Eq.~(\ref{eq:Markov}a), which is proportional
to $n_{\k_2}^{\up/\down}-n_{\k_1}^{\up/\down}$, describes a redistribution
of occupations of spin-up and spin-down states within a shell of
defined kinetic energy via the term proportional to 
$\delta(\omega_{\k_2}-\omega_{\k_1})$. For a parabolic band structure, this 
implies a redistribution between states with the same modulus $k$ of
the wave vector $\k$, while the total carrier spin remains unchanged.
If accompanied by a wave-vector dependent magnetic field like a
Rashba or the Dresselhaus field, this term leads to a 
D'yakonov-Perel'-type suppression of the spin dephasing.
Here, however, we do not consider any wave vector dependent field
and the system under investigation is isotropic in $\k$-space, so
that the first line in Eq.~(\ref{eq:Markov}a) has no influence on the
dynamics of the total spin. 
The second line in Eq.~(\ref{eq:Markov}a) describes a
spin-flip scattering from the spin-up band to the spin-down band
and vice versa. 
Since these bands are energetically split by $\hbar\we$
and a flip of carrier spin involves a corresponding flip of an
impurity spin in the opposite direction, which requires a magnetic 
(Zeeman) energy of $\hbar\wMn^\|$, the total magnetic energy released in a 
spin-flip process is $\pm\hbar(\we-\wMn^\|)$. Thus, 
$\delta\big[\omega_{\k_2}-\big(\omega_{\k_1}\pm (\we-\wMn^\|)\big)\big]$
ensures a conservation of the total single-particle energies in the 
Markov limit. 
It is noteworthy that, if the mean-field dynamics of the source terms
as in Eq.~(\ref{eq:bMF}) is not correctly taken into account,
other energetic shifts are obtained in the $\delta$-function,
which yields equations in the Markov limit that are not
consistent with the conservation of the single-particle energies \cite{kdep}.
Note also that the right-hand side of Eq.~(\ref{eq:Markov}a)
correctly deals with Pauli blocking effects.
Because the non-magnetic impurity scattering enters in the
equations of motion (\ref{eq:Markov}a) for the spin-up and spin-down
occupation only via the first line which plays no role in an isotropic system,
it has no influence on the spin dynamics in the Markov limit.

The first three lines in Eq.~(\ref{eq:Markov}b) 
for the perpendicular carrier spin component, which are proportional 
to $\delta$-functions, indicate an exponential decay 
of the perpendicular carrier spin component towards zero.
The last three lines describe a precession of the perpendicular carrier
spin component. The mean-field precession frequency $\we$ is 
renormalized by the carrier-impurity correlations. This renormalization 
originates from the imaginary part of the memory integral in 
Eq.~(\ref{eq:memint}). Besides the terms proportional to 
$\frac 1{\omega_{\k_2}-\big(\omega_{\k_1}\pm(\we-\wMn^\|)\big)}$, which
are also present when only the magnetic $s$-$d$ interaction is
taken into account \cite{FreqRenorm,*FreqRenorm_Erratum}, the non-magnetic impurity
scattering introduces another contribution which is a 
cross-term, i.e. it is absent when either the magnetic or
the non-magnetic impurity scattering is absent, which can be seen
from the fact that it is proportional to the product of $\Jsd$ and $J_0$.
In the quasi-continuous limit, the sum over $\k_2$ can be replaced by 
an integral over the spectral density of states.
In quasi-two-dimensional systems like quantum wells,
the spectral denstiy of states $D(\omega)=\frac{Am^*}{2\pi\hbar}$ is constant. 
Thus, the frequency renormalization can be integrated and yields
logarithmic divergences
\begin{align}
\label{eq:log}
&\sum_{\k_2}\frac 1{\omega_{\k_2}-\omega_{0}} =
\int\limits_0^{\omega_{BZ}} d\omega\, D(\omega) 
\frac 1{\omega-\omega_{0}}\nn&=
\frac{Am^*}{2\pi\hbar}\ln\left|\frac{\omega_{BZ}-\omega_0}{\omega_0}\right|.
\end{align}
at the poles $\omega_0=\omega_{\k_1}$ and 
$\omega_0=\omega_{\k_1}\pm(\we-\wMn^\|)$.
These logarithmic divergences are similar to the ones obtained in the discussion
of the Kondo-effect in metals with magnetic impurities \cite{Kondo}.
Despite the formal divergence, 
the summation over a non-singular carrier distribution always leads to a 
finite value of the precession frequency of the total carrier spin, since
the logarithm is integrable \cite{kdep}.
From Eq.~(\ref{eq:log}), one can see that the cut-off energy 
$\hbar\omega_{BZ}$, which corresponds to the width of the conduction band 
and is typically of the order of 1 eV, enters as a new model parameter in 
the theory and cannot be eliminated by assuming that $\omega_{BZ}\to \infty$,
since then the frequency renormalization also diverges.
As a consequence, the Markovian expression for the frequency renormalization
can only give an order-of-magnitude estimation and a more detailed 
treatment of the band structure is necessary if a quantitatively more 
accurate description is required.

For the special case of zero external magnetic field, vanishing
impurity magnetization and low carrier densities, 
Eqs.~(\ref{eq:Markov}) are equivalent to the simple rate equations
\begin{align}
\ddt \mbf s_{\k_1}=-\frac 1\tau
\mbf s_{\k_1},
\end{align}
where the values of the rates coincide with the Fermi's golden rule 
value. In two dimensions, one obtains\cite{Thurn:13_1}
\begin{align}
\label{eq:rateB0}
\frac 1{\tau^{2D}}=&
\frac{35}{12}\frac{\Jsd^2m^*}{\hbar^3}\frac{\NMn}V \frac 1d.
\end{align}

\subsection{Correlation energy}
In Eqs.~(\ref{eq:Qint}) to (\ref{eq:memint}), 
Markovian expressions for the carrier-impurity correlations
are derived as functionals of the carrier 
and impurity variables.
Using these expressions, it is straightforward to also obtain  
analytic expressions for the carrier-impurity correlation energies
as functionals of the carrier spins and occupations \cite{kdep}.
Splitting the averages over the magnetic and non-magnetic carrier-impurity
interactions into mean-field and correlated contributions
\begin{subequations}
\begin{align}
\langle \Hsd\rangle=&\langle \Hsd^\t{MF} \rangle
+\langle \Hsd^\t{cor}\rangle,\\
\langle \Himp\rangle=&\langle \Himp^\t{MF} \rangle
+\langle \Himp^\t{cor}\rangle,\\
\langle \Hsd^\t{MF} \rangle=&\frac{\Jsd \NMn}{V} \sum_\k
\langle \mbf S\rangle\cdot \mbf s_{\k}\\
\langle \Hsd^\t{cor}\rangle=&\frac{\Jsd \NMn }{V^2} \sum_{\k,\k'}
\sum_i Q_{i\k}^{i\k'}\\
\langle \Himp^\t{MF}\rangle=&
\frac{J_0 \Nimp }{V} \sum_\k n_\k,\\
\langle \Himp^\t{cor}\rangle=& \frac{J_0 \Nimp }{V^2} \sum_{\k,\k'}
\bar{C}_{\phantom{0}\k}^{0\k'},
\end{align}
\end{subequations}
one obtains in the Markov limit
\begin{subequations}
\label{eq:cor_en}
\begin{align}
\langle \Hsd^\t{cor}\rangle=&-\frac{\Jsd\NMn}{V^2}\sum_{\k_1\k_2}\bigg\{
\frac{\frac 12\Jsd\langle {S^\|}^2\rangle n_{\k_1}
+2J_0\langle S^\|\rangle s_{\k_1}^\|}{\omega_{\k_2}-\omega_{\k_1}}\nn&
+\frac{\Jsd\big(\langle{S^\perp}\rangle-\frac 12\langle S^\|\rangle\big)
 (1-n_{\k_2}^\down)n_{\k_1}^\up}
{\omega_{\k_2}-\big(\omega_{\k_1}+(\we-\wMn^\|)\big)}\nn&
+\frac{\Jsd\big(\langle{S^\perp}\rangle+\frac 12\langle S^\|\rangle\big)
 (1-n_{\k_2}^\up)n_{\k_1}^\down}
{\omega_{\k_2}-\big(\omega_{\k_1}-(\we-\wMn^\|)\big)}
\bigg\},\\
\langle \Himp^\t{cor}\rangle=&
-2\frac{J_0\Nimp}{V^2}\sum_{\k_1\k_2}\frac{J_0n_{\k_1}+
\Jsd\langle S^\|\rangle s_{\k_1}^\|}{\omega_{\k_2}-\omega_{\k_1}}.
\end{align}
\end{subequations}

Eqs.~(\ref{eq:cor_en}) have the same poles as Eq.~(\ref{eq:Markov}b) for
the frequency renormalization and, thus, also contain formally logarithmic
divergences in two-dimensional systems.

\section{Results}
\begin{figure*}
\includegraphics[width=\textwidth]{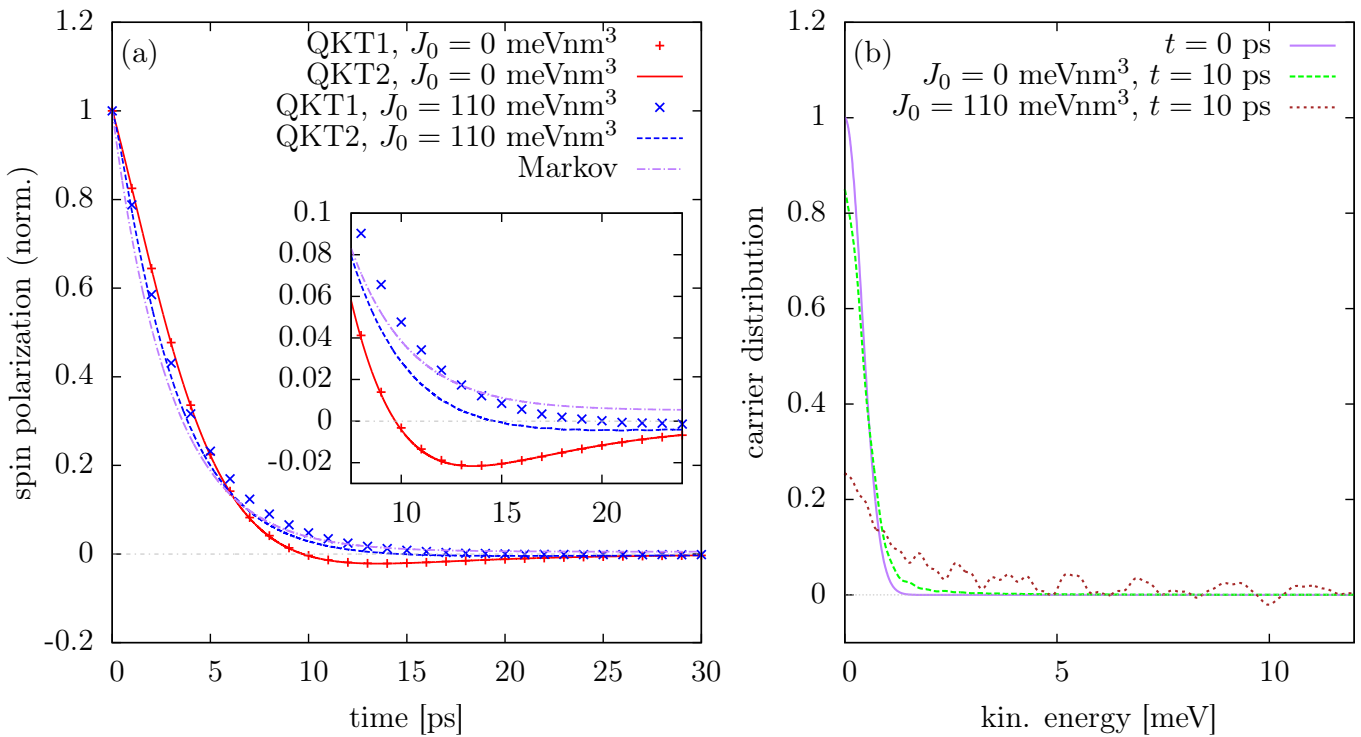}
\caption{(a): Time evolution of the carrier spin for zero magnetic field
with ($J=110$ meVnm$^3$) and without ($J=0$) non-magnetic impurity scattering.
QKT1 (points) denotes the results according to the full quantum 
kinetic equations~(\ref{eq:corful}) while QKT2 (lines)
describes the results of the reduced
set of equations (\ref{eq:eom_sum}). The purple dash-dotted line shows
the results of the Markovian equations (\ref{eq:Markov}), which is
independent of non-magnetic impurity scattering.
The inset shows a magnification of the region where the quantum kinetic
theory for $J_0=0$ predicts a non-monotonic behavior.
(b) Occupation of carrier states at $t=0$ and $t=10$ ps for the 
calculations with and without non-magnetic impurity scattering.
\label{fig:B0}}
\end{figure*}

After having derived the quantum kinetic equations for the description of
the spin dynamics in DMS including magnetic and non-magnetic scattering
and having obtained rate-type Markovian equations, 
we now present results of numerical simulations. Here, we focus 
on the case of a 4-nm-wide Cd$_{0.93}$Mn$_{0.07}$Te quantum well.
For this material, the magnetic coupling constant is $\Jsd=-15$ meVnm$^3$
($N_0\Jsd=-220$ meV) \cite{intro}, while the non-magnetic coupling constant 
is approximately $J_0=110$ meVnm$^3$ ($N_0J_0=1.6$ eV) 
\cite{CdTe_offset_Nawrocki}, where $N_0$ is the number of unit
cells per unit volume. Furthermore, we use a conduction band effective
mass of $m^*=0.1\, m_0$ and assume that the impurity magnetization is
described by a thermal distribution at a temperature of $T=2$ K
and the g-factors of the conduction band carriers and Mn impurities 
are $g_e=-1.77$ and $g_\t{Mn}=2$, respectively \cite{SPIE}. If not stated
otherwise, we choose a value of 40 meV for the cut-off energy 
$\hbar\omega_{BZ}$ in the numerical calculations and we consider only
Mn ions as sources of non-magnetic impurity scattering, i. e. $\Nimp=\NMn$.
As initial value for the carrier distribution, we use a Gaussian distribution
centered at the band edge of the spin-up band with standard deviation of
$E_s=0.4$ meV, which corresponds to an excitation with a circularly 
polarized light pulse with full width at half maximum (FWHM) pulse duration
of about 350 fs.

We first discuss the spin dynamics in the conduction band of
a Cd$_{0.93}$Mn$_{0.07}$Te quantum well for zero magnetic field 
with a focus on the impact of non-magnetic impurity scattering
on the spin dynamics and investigate the
redistribution of carriers in $\k$-space 
as well as the build-up of correlation energy.
Then, we study the spin dynamics in the valence band 
in a simplified model.
Finally, we investigate the spin dynamics in the
presence of an external magnetic field parallel and perpendicular
to the carrier spin polarization and discuss, in the latter case,
how the non-magnetic impurity scattering affects the 
carrier spin precession frequencies.

\subsection{Zero magnetic field}
Figure \ref{fig:B0}(a) shows the time evolution of an initially polarized
carrier spin in a Cd$_{0.93}$Mn$_{0.07}$Te quantum well for vanishing 
magnetic field. 
The Markovian equations (\ref{eq:Markov}) predict a simple exponential 
decay of the carrier spin, which is transferred to the impurities.
Note that due to $\NMn\gg N_e$, the asymptotic value of the carrier spin 
for long times $t$ is close to zero, 
since the impurities act as a spin bath. 
If only the magnetic spin-flip scattering is accounted for ($J_0=0$),
the time evolution according to the quantum kinetic theory is non-monotonic
and shows an overshoot below the asymptotic value. 
These non-Markovian effects are strongly suppressed in the
calculations including non-magnetic impurity scattering ($J_0=110$ meVnm$^3$)
and the time evolution of the total spin follows the Markovian dynamics 
more closely. An exponential fit to the dynamics of the full quantum kinetic 
theory yields an effective spin transfer rate about $15\%$ smaller 
than the Markovian rate in Eq.~(\ref{eq:rateB0}).

\begin{figure}
\includegraphics{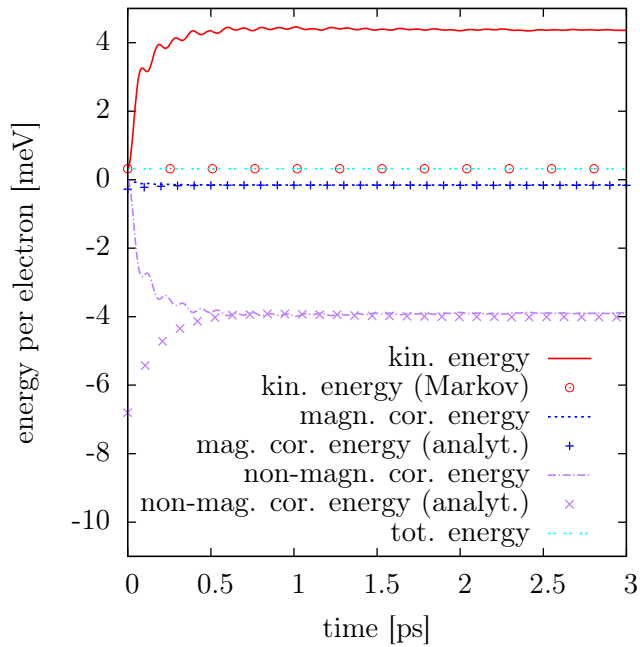}
\caption{Kinetic energy (red line), magnetic correlation energy (blue line),
non-magnetic correlation energy (purple line) and total energy (cyan line)
per electron for the quantum kinetic calculation shown in Fig.~\ref{fig:B0} 
with $J_0=110$ meVnm$^3$. 
The red circles show the kinetic energy obtained from 
the Markovian calculation in Fig.~\ref{fig:B0}.
The pluses and crosses depict the results according to the 
analytic Markovian expressions for the correlation energies in 
Eqs.~(\ref{eq:cor_en}) evaluated using the carrier distribution 
of the quantum kinetic calculation at selected time steps.
\label{fig:cor_en}
}
\end{figure}

Interestingly, while the full quantum kinetic equations
(\ref{eq:corful}) yield identical results as the reduced set of
equations (\ref{eq:eom_sum}) in the case without non-magnetic impurity 
scattering, deviations between both approaches can be clearly seen 
when the non-magnetic impurity scattering is taken into account.

In order to understand the suppression of the non-monotonic features in
the spin dynamics with non-magnetic impurity scattering,
it is useful to recapitulate the findings of Ref.~\onlinecite{Proceedings}, 
where the origin of the non-Markovian behavior of the spin dynamics in absence
of non-magnetic impurity scattering was discussed:  
It was found that the depth of the memory induced by the correlations
is of the order of the inverse energetic distance of the 
carrier state under consideration to the band edge times $\hbar$. 
Memory effects 
become insignificant if the kinetic energy of the carrier $\hbar\omega_{\k_1}$
is much higher than the energy scale of the carrier-impurity spin 
transfer rate $\frac\hbar\tau$. For the parameters used in the simulations,
one obtains from Eq.~(\ref{eq:rateB0}) a value of $\tau^{2D}=2.97$~ps and
therefore $\frac\hbar\tau\approx 0.22$~meV. 
Figure \ref{fig:B0}(b) shows the redistribution of carriers in the
calculations with and without non-magnetic impurity scattering.
One can clearly see that, while without non-magnetic impurity scattering
the carrier distribution at $t=10$ ps is only slightly broadened, 
including the non-magnetic impurity scatterings leads to a
drastic redistribution of carriers to states many meV away from the initial 
distribution. For these states, the memory is very short compared with the
spin relaxation time and therefore the Markovian approximation is
justified.

The redistribution of carriers to states several meV away from the band edge 
raises questions about the conservation of energy, since for zero magnetic field
the mean-field energy of the system is comprised of only the kinetic
energy of the carriers. In the quantum kinetic calculations, however,
we also consider the carrier-impurity correlations which introduce
correlation energies that are not captured in a simple single-particle picture.
The different contributions to the total energy over the course of time
for the simulations presented in Fig. \ref{fig:B0} are shown 
in Fig. \ref{fig:cor_en}. There, it is shown that the average kinetic
energy per electron increases from the initial value of the order of
the width of the initial carrier distribution to a much larger value
of about $4$ meV on a timescale of about $0.5$ ps.
This energy is mostly provided by a decrease of non-magnetic correlation
energy from zero to a negative value. 
The magnetic correlation energy is comparatively small since the 
magnetic coupling constant $\Jsd$ is about one order of magnitude
smaller than the non-magnetic coupling constant $J_0$.
The pluses and crosses in Fig. \ref{fig:cor_en} show the results of the 
analytic expressions~(\ref{eq:cor_en}) for the correlation
energies evaluated using the carrier distributions of the full quantum kinetic
theory in the respective time steps. 
The analytic results are found to coincide with the values extracted
from the quantum kinetic theory after the first $0.5$ ps.
Even though the analytic expressions for the correlation energies are obtained 
within the Markovian description, it should be noted that in the Markovian 
equations of motion (\ref{eq:Markov}) for the spins and occupations 
only single-particle energies are considered for evaluating the 
energy balance.
As in our case the single particle energies comprise only the 
kinetic energies of
the carriers, the latter are constant in the Markovian description 
in sharp contrast
to the quantum kinetic treatment.

Note also that the total energy comprised of 
single-particle and correlation energies remains constant in the 
quantum kinetic simulations, which provides a further test for the numerics.  

\subsection{Valence band} 
The fact that in the conduction band of a Cd$_{1-x}$Mn$_x$Te quantum
well the non-magnetic scattering at the impurities suppresses the
characteristic non-monotonic features of genuine quantum kinetic 
behavior raises the question whether this statement is true in general 
and non-Markovian effects always only change the spin dynamics quantitatively.
In this section, we provide an example of a situation where the
non-Markovian features are not suppressed due to impurity scattering.

\begin{figure}
\includegraphics{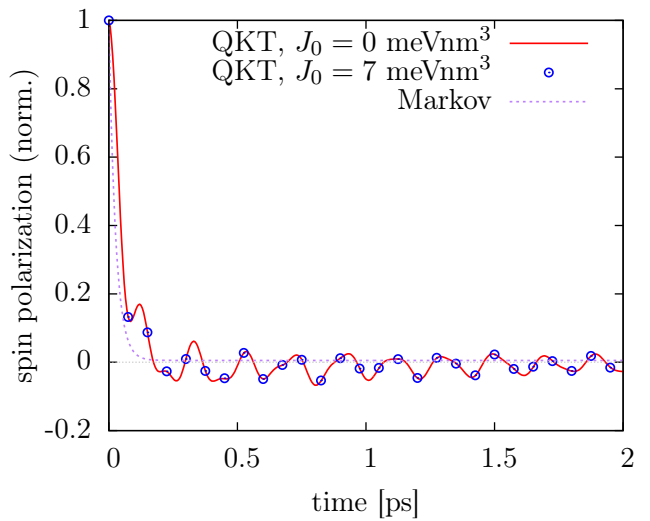}
\caption{Spin dynamics in a degenerate valence band of a 
Cd$_{0.93}$Mn$_{0.07}$Te quantum well with and without accounting
for non-magnetic impurity scattering.}
\label{fig:val}
\end{figure}

We consider now the valence band of a Cd$_{1-x}$Mn$_x$Te quantum well.
The details of the valence band structure in a quantum well are
influenced by, e.g., spin-orbit coupling, strain or the shape of
the confinement potential. A realistic description of the band structure
is beyond the scope of this article. Instead, we perform a model study, where
we assume that heavy-hole and light-hole bands are degenerate. In this
case, we can use the quantum kinetic theory derived for the conduction 
band and take the material parameters for the heavy holes.
The magnetic coupling constant in the valence band is $\Jpd=60$ meVnm$^3$
\cite{intro} and the heavy-hole mass is $m_h=0.7 m_0$ \cite{CdTe_meff}.
The difference of the band gaps between CdTe and zinc-blende MnTe
of about $1.6$ eV is split into the conduction and valence band offsets
by a ratio of 14:1 \cite{CdTe_offset}. Thus, one obtains a 
value for the non-magnetic coupling constant in the valence band 
of about $J_0=7$ meVnm$^3$.
The results of the quantum kinetic simulations for these parameters are
shown in Fig.~\ref{fig:val}.

\begin{figure*}
\includegraphics{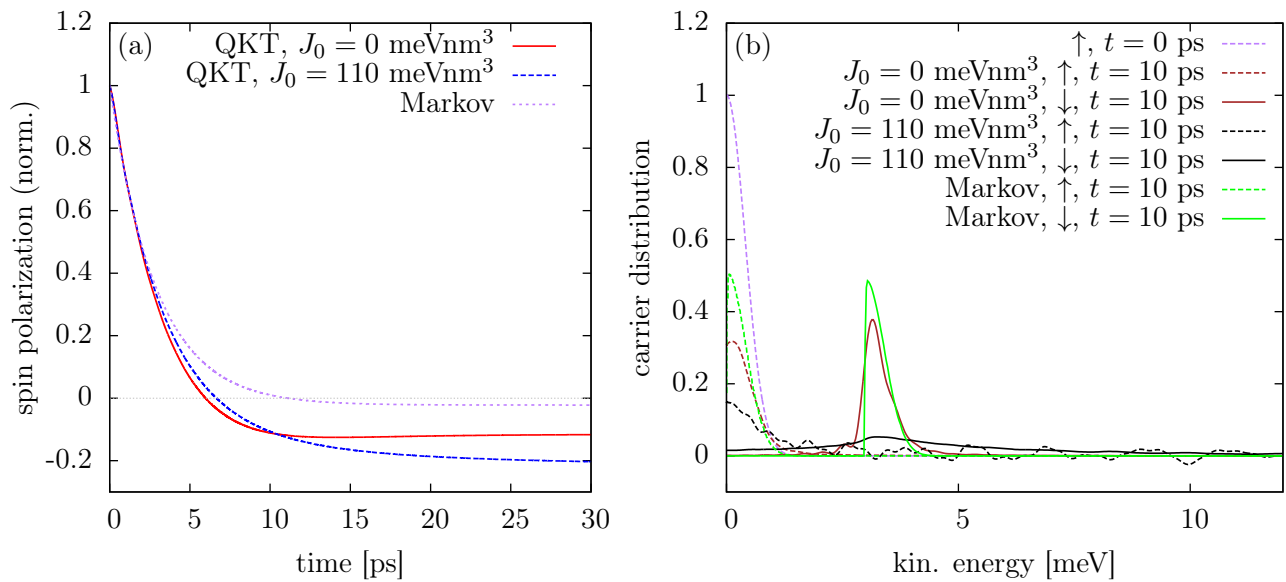}
\caption{(a): Time evolution of the carrier spin polarization parallel to 
an external magnetic field ($B=100$ mT).
(b): Spin-up ($\up$) and spin-down occupations ($\down$) at $t=0$ and $t=10$ ps.
\label{fig:bz1}}
\end{figure*}

In comparison with the conduction band, the 4 times larger 
magnetic coupling constant in the valence band leads to much stronger
non-Markovian effects. In particular, one finds not a single overshoot,
but pronounced oscillations of the spin polarization about its 
asymptotic value. 
In Fig.~\ref{fig:val}, the calculations with and without accounting
for non-magnetic impurity scattering yield practically identical results.
Thus, due to the fact that in the valence band the non-magnetic coupling
constant is much smaller than the magnetic coupling constant,
no suppression of non-Markovian effects in the spin dynamics is 
observed.

%
%
%

\subsection{Finite magnetic field: Faraday configuration}
Next, we investigate the effects of non-magnetic impurity scattering
on the spin dynamics in DMS in the presence of an external magnetic
field. In this section, we study the case in which the external field and
the initial carrier spins are parallel, which is known as the Faraday 
configuration. This case has also been considered in 
Ref.~\onlinecite{SPIE}, but without accounting for
 non-magnetic impurity scattering. 

In Fig.~\ref{fig:bz1}(a) the time evolution of the carrier spin polarization
parallel to an external magnetic field $B=100$ mT is shown. 
Note that the non-monotonic behavior that can be seen in the 
case without an external magnetic field is suppressed for finite 
external fields even if the non-magnetic scattering is disregarded.
The most striking feature in the time evolution of the carrier spin polarization
is that the Markovian result and the quantum kinetic simulations predict
very different asymptotic values of the spin polarization 
at long times $t$.

As discussed in Ref.~\onlinecite{SPIE}, the different stationary values are 
related to a broadening of the distribution of scattered carriers in the
spin-down band, which is shown in Fig.~\ref{fig:bz1}(b). 
Note that the broadening of the carrier distribution is not 
primarily an effect of energy-time uncertainty, which is commonly found
in quantum kinetic studies \cite{phonon_replica_theo,phonon_replica_exp}, 
since the width of the 
distribution does not shrink significantly over the course of time
\cite{SPIE}. Rather, it is a consequence of the build-up of correlation energy 
which enables deviations from the conservation of the single-particle energies
in spin-flip scattering processes.

In the Markov limit, the stationary value 
is obtained when a balance between scattering 
from the spin-up to the spin-down band and vice versa is reached.
In the quantum kinetic calculations, the distribution of the scattered
carriers is broadened, so that also spin-down states
below the threshold $\hbar\we-\hbar\wMn^\|$ are occupied, whose back-scattering
is suppressed since there are no states in the spin-up band with the 
matching single-particle energies.
If additionally the non-magnetic impurity scattering is taken into account, 
the scattered impurity distribution is even broader and more 
spin-down states with kinetic energies below $\hbar\we-\hbar\wMn^\|$ are 
occupied, so that the back-scattering is more strongly suppressed and the
deviation of the asymptotic value of the spin polarization from the
Markovian value is even larger.

\subsection{Finite magnetic field: Voigt configuration}
\begin{figure*}
\includegraphics{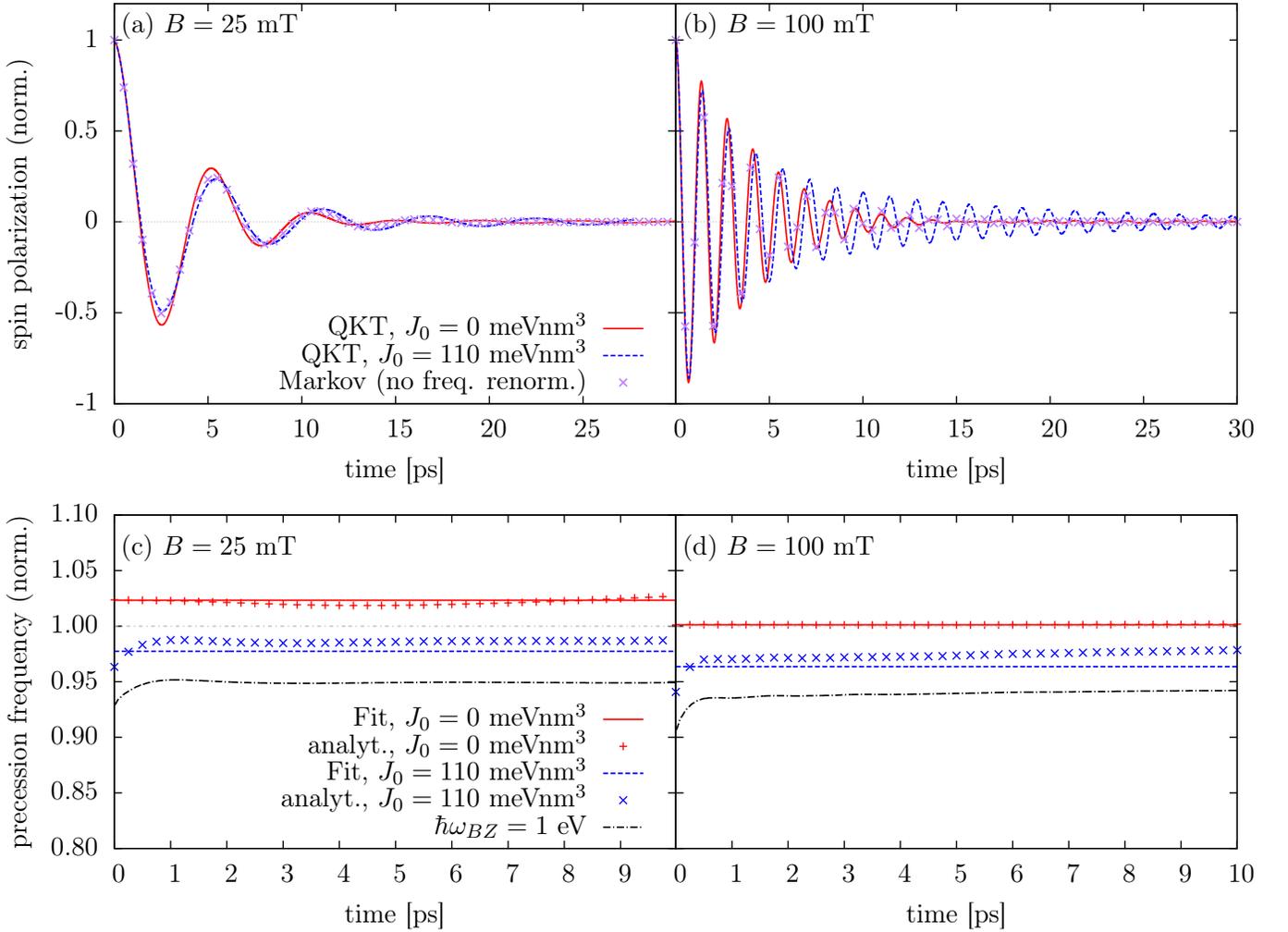}
\caption{(a) and (b): Time evolution of the carrier spin polarization 
for $B=25$ mT (a) and $B=100$ mT (b) using the quantum kinetic 
equations~(\ref{eq:corful}) 
and the Markovian equations Eq.~(\ref{eq:Markov}b), where the terms
responsible for the frequency renormalization  in the Markovian equations
have been dropped.
The precession frequency normalized
with respect to its mean-field value $\we$ is shown in (c) and (d) 
using a fit of an exponentially decaying cosine 
to the quantum kinetic results 
and the analytic expressions obtained from Eq.~(\ref{eq:Markov}b)
and the occupations from the quantum kinetic calculations.
The black dash-dotted lines in (c) and (d) show the analytic results
for a cut-off energy of $\hbar\omega_{BZ}=1$ eV.
\label{fig:bx}}
\end{figure*}

The situation in which an external magnetic field and the optically induced
carrier spin polarization are perpendicular to each other
is usually referred to as
the Voigt configuration and is the subject of this section.
In this situation, the carrier spin precesses about the 
effective magnetic field $\bwe$ due to the external field and 
the impurity magnetization.
As shown in Ref.~\onlinecite{FreqRenorm,*FreqRenorm_Erratum}, where the non-magnetic impurity
scattering was disregarded, the carrier-impurity correlations 
are responsible for a renormalization of the precession frequency.
There, it was also shown that the strength of this renormalization depends
on the details of the carrier distribution and the strength of the 
effective field $\bwe$.

In Fig. \ref{fig:bx}(a), we present simulations of the spin dynamics in
a DMS in Voigt geometry for an external magnetic field of $B=25$ mT,
which corresponds to a situation with $|\langle \mbf S\rangle|\approx 0.05$, 
where the magnetic-correlation-induced frequency renormalization
according to Ref.~\onlinecite{FreqRenorm,*FreqRenorm_Erratum} is particularly strong.
Simulations with ($J_0=110$ meVnm$^3$) and without ($J_0=0$) 
accounting for the non-magnetic impurity scattering are compared to
Markovian calculations based on Eqs.~(\ref{eq:Markov}). Note that
for the Markovian calculation shown in Fig. \ref{fig:bx} the 
frequency renormalization was not taken into account.
The results of all simulations shown in Fig. \ref{fig:bx}(a)
are very similar and follow closely the form of an exponentially damped 
cosine. Note that at long times, the phases of the oscillations 
of the calculations accounting for non-magnetic impurity scattering
matches the Markovian calculation without frequency renormalization, while
accounting only for magnetic spin-flip scattering leads to oscillations
with a slightly higher frequency. 

The frequency renormalization for the simulations shown in Fig. \ref{fig:bx}(a)
is presented in Fig. \ref{fig:bx}(c), where an exponentially decaying
cosine is fit to the quantum kinetic results and, for comparison,
the total precession frequency including the correlation-induced 
renormalization in the Markovian description in Eq.~(\ref{eq:Markov}b) 
evaluated using the spin-up and spin-down occupations of the quantum kinetic
simulations is depicted. 
Due to the time evolution of the occupations, also the renormalization
predicted by Eq.~(\ref{eq:Markov}b) becomes a function of time,
which, however, is for all times close to the constant extracted by
fitting the quantum kinetic result.
The calculations without non-magnetic impurity scattering predict 
an increase of the carrier spin precession frequency of about $2-3\%$ 
with respect to the mean-field value $\we$, which is consistent with the
findings of Ref.~\onlinecite{FreqRenorm,*FreqRenorm_Erratum}. 
On the other hand, the contribution from the non-magnetic carrier-impurity
correlations leads to a decrease of the precession frequency which partially
cancels the contribution from the magnetic correlations.

In Figs. \ref{fig:bx}(b) and \ref{fig:bx}(d), the time evolution of the
carrier spin polarization and the frequency renormalization are shown 
for an external magnetic field of $B=100$ mT. 
In this case, the envelope of the spin polarization 
decays only exponentially for the calculations without non-magnetic impurity 
scattering. For $J_0=110$ meVnm$^{3}$, the spin polarization follows
the exponential decay of the simulation with $J_0=0$ only up to about 5 ps. 
After that, it decays much slower, which is a new non-Markovian effect 
that is absent if the non-magnetic impurity scattering is disregarded.
As can be seen in Fig.~\ref{fig:bx}(d),
the frequency renormalization 
due to the magnetic interaction alone is almost zero.
Nevertheless, when the non-magnetic carrier-impurity correlations are
taken into account, the precession frequency shows a decrease of 
about $2-3\%$. Thus, in contrast to the correlation-induced renormalization
in absence of non-magnetic scattering where the renormalization 
is only observable for a very narrow set of initial conditions
\cite{FreqRenorm,*FreqRenorm_Erratum}, including the non-magnetic 
carrier-impurity interaction results in a significant renormalization for
a much broader set of excitation conditions.

It is noteworthy that the frequency renormalization in the quantum kinetic
calculations is well reproduced by the Markovian expression 
in Eq.~(\ref{eq:Markov}b). The numerical demands of the full 
quantum kinetic equations require a restriction of the conduction band 
width $\hbar\omega_{BZ}$ used in the calculations to a few tens of meV. 
However, in realistic band structures, the band widths are of the order of
eV. In order to give an order-of-magnitude estimation of the 
frequency renormalization for such band widths, we present in 
Figs.~\ref{fig:bx}(c) and \ref{fig:bx}(d) also the results of the Markovian
expression for the frequency renormalizations using the value of
$\hbar\omega_{BZ}=1$ eV together with the occupations obtained in the
quantum kinetic calculations for $\hbar\omega_{BZ}=40$ meV. This estimation
yields a renormalization of the precession frequencies due to the
combined effects of magnetic and non-magnetic scattering of about
$5-7\%$. A quantitatively more accurate description requires a more detailed
treatment of the band structure, which is beyond the scope of this article.

Note also that the frequency renormalization due to the non-magnetic 
carrier-impurity correlations is dominated by a cross-term proportional 
to $\Jsd J_0$ [cf. fourth line in Eq.~(\ref{eq:Markov}b)].
Thus, the sign of the frequency renormalization depends on the 
relative signs of the coupling constants $\Jsd$ and $J_0$. In principle, this
allows a determination of the sign of the magnetic coupling constant
$\Jsd$ which cannot be obtained directly, e.g., by measuring the
giant Zeeman splitting of excitons \cite{intro}.

\section{Conclusion}
We have investigated the influence of non-magnetic impurity scattering
at Mn impurities on the spin dynamics 
in Cd$_{1-x}$Mn$_x$Te diluted magnetic semiconductors.
To this end, we have developed a quantum kinetic theory taking 
the magnetic and non-magnetic carrier-impurity correlations into account.
The Markov limit of the quantum kinetic equation is derived in order to
distinguish the Markovian dynamics from genuine quantum kinetic effects.

In contrast to earlier studies \cite{Thurn:13_1, Thurn:13_2, PESC, SPIE}
in which only the magnetic contribution to the carrier-impurity interaction
has been considered, some non-Markovian effects, such as a non-monotonic 
spin transfer between carriers and impurities, are strongly suppressed
in the case of the conduction band of a Cd$_{1-x}$Mn$_x$Te quantum well,
while other features stemming from non-Markovian dynamics are enhanced,
such as the large finite stationary value of the spin polarization 
in a magnetic field reached at long times.
The reason for the suppression in the former case
 is that the non-magnetic impurity scattering
leads to a strong redistribution of carriers in $\k$-space away from
the states at $\k=0$. Since memory effects are particularly strong
for carriers in proximity to the band edge \cite{PESC}, this redistribution
leads to spin dynamics that are well described by Markovian rate equations.
The redistribution of the carriers implies an increase of their kinetic energies
which is provided by a build-up of (negative) carrier-impurity 
correlation energy and which cannot be described by a mean-field or 
semiclassical approximation. 
We also provide analytic expressions for the
correlation energies in the form of functionals of the spin-up and
spin-down carrier occupations. Numerical calculations confirm that
these expressions indeed describe the correlation energies obtained from 
the full quantum kinetic theory very well.

Even though doping with magnetic impurities unavoidably also 
provides a contribution to non-magnetic impurity scattering, there 
can still be situations where the latter is too weak to influence the
spin dynamics and to suppress otherwise visible non-Markovian effects.
This is substantiated by a model study of a Cd$_{1-x}$Mn$_x$Te quantum well 
with degenerate valence bands, where the spin polarization exhibits a 
non-monotonic time dependence in the form of oscillations,
while the Markovian treatment predicts a simple exponential decay.
Further investigations using a more realistic valence
band structure are needed in order to make more precise predictions about
possible non-Markovian features in the hole spin dynamics in DMS.

%
In the presence of an external magnetic field parallel to the initial
carrier spin (Faraday geometry), earlier studies \cite{SPIE} that
did not consider non-magnetic impurity scattering predicted that
the asymptotic value of the carrier spins in the conduction band 
of a DMS quantum well at long times $t$ are significantly
different in quantum kinetic and Markovian calculations. This was
attributed to a broadening of the distribution of the scattered electrons 
due to the build-up of strong carrier-impurity correlations, which, 
because of the correlation energy, leads to a non-conservation of 
single-particle energies. The broadening results in an occupation of states
by electrons whose back-scattering to the original band is
strongly suppressed due to the lack of states with matching single-particle
energies. This induces a bias between spin-flip scattering from the
spin-up to the spin-down subband and vice versa. 
In the presence of a strong non-magnetic carrer-impurity interaction, the
correlation energy becomes much larger and with it also the broadening
of the scattered carrier distribution and the deviations of the
asymptotic value of the carrier spin polarization from its value 
obtained in Markovian calculations.

In the Voigt geometry, where the initial carrier spin polarization is
perpendicular to the external field, the carrier spin precesses about
the effective magnetic field comprised of the external field and the
mean field due to the impurity magnetization. There, the carrier-impurity
correlations lead to a renormalization of the spin precession frequencies.
An analytic expression for this renormalization is presented and it is
found to be of a similar form as the expression for the correlation 
energies. The non-magnetic carrier-impurity interaction influences
the frequency renormalization via a cross-term which vanishes if either
the magnetic or the non-magnetic carrier-impurity interaction is neglected.
In the case of the conduction band of Cd$_{1-x}$Mn$_x$Te, the magnetic 
and non-magnetic contributions to the frequency renormalization have
opposite signs. 
A measurement of the frequency renormalization can therefore indicate
the sign of the exchange interaction.
For magnetic fields at which the renormalization 
due to the magnetic correlations is particularly strong, the magnetic
and non-magnetic contributions almost cancel each other. However,
in most situations, the purely magnetic contribution is relatively weak
\cite{FreqRenorm,*FreqRenorm_Erratum}, so that the cross-term dominates the total 
frequency renormalization. The order of magnitude of the frequency 
renormalization for the cases considered here is about a few percent
of the mean-field precession frequency. 

To summarize, the influence of the non-magnetic impurity scattering on the
spin dynamics in DMS is two-fold: First, it leads to a significant
redistribution 
of carriers in $\k$-space, which facilitates the suppression of some
non-Markovian effects in certain situations. Second, it causes a
formation of strong many-body correlations between carriers and
impurities, which result in large correlation energies and 
a significant renormalization of the carrier spin precession frequency.

\begin{acknowledgments}
We gratefully acknowledge the financial support from the Universidad de
Buenos Aires, project UBACyT 2014-2017 No. 20020130100514BA,
and from CONICET, project PIP 11220110100091.
\end{acknowledgments}

\appendix
\section{Reduced set of equations of motions\label{app:neweqs}}
The equations of motions for the variables defined in 
Eq.~(\ref{eq:newvars}) are:

\begin{widetext}
\begin{subequations}
\label{eq:eom_sum}
\begin{align}
\ddt \langle S^l\rangle=&\sum_{im}\epsilon_{lim}\wMn^i \langle S^m\rangle
+\frac{\Jsd}V \sum_{\k\k'}\sum_{im} \epsilon_{lim} \Re\{Q_{m\k}^{i\k'} \},\\
\ddt n_{\k_1}=&\frac{\Jsd\NMn}{\hbar V^2} \sum_{\k}\sum_i2\Im\{Q_{i\k_1}^{i\k}\}
+\frac{J_0\Nimp}{\hbar V^2}\sum_\k 2\Im\{\bar{C}_{\phantom{0}\k_1}^{0\k}\},\\
\ddt s^l_{k_1}=&\sum_{ij}\epsilon_{lij}\we^i s_{\k_1}^j+
\frac{\Jsd\NMn}{\hbar V^2} \sum_{\k} \big[
\frac 12 \Im\{Q_{l\k_1}^{0\k}\}+\sum_{ij}
\epsilon_{lij}\Re\{Q_{i\k_1}^{j\k}\}\big]
+\frac{J_0\Nimp}{\hbar V^2}\sum_\k 2\Im\{\bar{C}_{\phantom{l}\k_1}^{l\k}\},\\
\ddt Q_{l\k_1}^{0\k_2}=&
-i(\omega_{\k_2}-\omega_{\k_1})Q_{l\k_1}^{0\k_2}
+\sum_{ii'}\epsilon_{l ii'}\wMn^i Q_{i'\k_1}^{0 \k_2}
+\frac i\hbar {b_{l\k_1}^{0\k_2}}^I
+\frac i\hbar {b_{l\k_1}^{0\k_2}}^\t{imp},\\
\ddt Q_{l\k_1}^{j\k_2}=&
-i(\omega_{\k_2}-\omega_{\k_1})Q_{l\k_1}^{j\k_2}
+\sum_{ii'}\epsilon_{j ii'}\we^i Q_{l\k_1}^{i'\k_2}
+\sum_{ii'}\epsilon_{l ii'}\wMn^i Q_{i'\k_1}^{j\k_2}
+\frac i\hbar {b_{l\k_1}^{j\k_2}}^I 
+\frac i\hbar {b_{l\k_1}^{j\k_2}}^\t{imp},\\
\ddt \bar{C}_{\phantom{0}\k_1}^{0\k_2}=&
-i(\omega_{\k_2}-\omega_{\k_1})\bar{C}_{\phantom{0}\k_1}^{0\k_2}
+\frac i\hbar {c_{\phantom{0}\k_1}^{0\k_2}}^I
+\frac i\hbar {c_{\phantom{0}\k_1}^{0\k_2}}^\t{sd},\\
\ddt \bar{C}_{\phantom{j}\k_1}^{j\k_2}=&
-i(\omega_{\k_2}-\omega_{\k_1})\bar{C}_{\phantom{j}\k_1}^{j\k_2}
+\sum_{ii'}\epsilon_{j ii'}\we^i \bar{C}_{\phantom{i'}\k_1}^{i'\k_2}
+\frac i\hbar {c_{\phantom{j}\k_1}^{j\k_2}}^I
+\frac i\hbar {c_{\phantom{j}\k_1}^{j\k_2}}^\t{sd},
\end{align}
with
\begin{align}
{b_{l\k_1}^{0\k_2}}^I=&
\Jsd\sum_i\bigg[\Re\{\langle S^i S^l\rangle\}
(s_{\k_2}^i-s_{\k_1}^i)
+i\sum_m\epsilon_{ilm}\frac{\langle S^m\rangle}2
\Big((1-n_{\k_1})s^i_{\k_2}+(1-n_{\k_2})s^i_{\k_1}\Big)
+\langle S^i\rangle(s_{\k_1}^l s_{\k_2}^i-s_{\k_1}^is_{\k_2}^l)\bigg],\\
{b_{l\k_1}^{j\k_2}}^I=&\Jsd\sum_i\bigg[
\Re\{\langle S^iS^l\rangle\}\Big[
\delta_{ij}\big(\frac{n_{\k_2}}4 -\frac{n_{\k_1}}4\big)
+\frac i2\epsilon_{ijk} (s^k_{\k_1}+s^k_{\k_2})\Big]+
\frac i2\sum_m\epsilon_{ilm}\langle S^m\rangle\Big[
\delta_{ij}\frac{n_{\k_1}+n_{\k_2}-n_{\k_1}n_{\k_2}}4 \nn&
+\delta_{ij}\mbf s_{\k_1}\cdot \mbf s_{\k_2}
-(s^i_{\k_1}s^j_{\k_2}+s^i_{\k_2}s^j_{\k_1})
+\frac i2\epsilon_{ijk}\big( (1-n_{\k_1})s_{\k_2}^k
-(1-n_{\k_2})s_{\k_1}^k\big)\Big]\bigg],\\
{b_{l\k_1}^{0\k_2}}^\t{imp}=&J_0\langle S^l\rangle (n_{\k_2}-n_{\k_1}),\\
{b_{l\k_1}^{j\k_2}}^\t{imp}=&J_0\langle S^l\rangle(s^j_{\k_2}-s^j_{\k_1}),\\
{c_{\phantom{0}\k_1}^{0\k_2}}^I=&J_0(n_{\k_2}-n_{\k_1}),\\
{c_{\phantom{j}\k_1}^{j\k_2}}^I=&J_0(s^j_{\k_2}-s^j_{\k_1}),\\
{c_{\phantom{0}\k_1}^{0\k_2}}^\t{sd}=&
\Jsd\langle S^i\rangle (s^i_{\k_2}-s^i_{\k_1}),\\
{c_{\phantom{j}\k_1}^{j\k_2}}^\t{sd}=&
\Jsd\Big[\frac 14\langle S^j\rangle(n_{\k_2}-n_{\k_1})
+\frac i2\epsilon_{ijk}\langle S^i\rangle(s^k_{\k_2}+s^k_{\k_1})\Big].
\end{align}
\end{subequations}
\end{widetext}

\bibliography{alle}

\end{document}